\documentclass[conference]{IEEEtran}
\IEEEoverridecommandlockouts
\usepackage{cite}
\usepackage{amsmath,amssymb,amsfonts}
\usepackage{algorithmic}
\usepackage{graphicx}
\usepackage{textcomp}
\usepackage{xcolor}
\usepackage{float}
\usepackage{multirow}
\usepackage{mdframed}
\usepackage{xcolor}
\usepackage{tabularx}
\usepackage[most]{tcolorbox}
\usepackage{fontawesome5}   
\usepackage{tcolorbox}
\usepackage{lipsum}         

\usepackage{listings}
\definecolor{codegreen}{rgb}{0,0.6,0}
\definecolor{codegray}{rgb}{0.5,0.5,0.5}
\definecolor{codepurple}{rgb}{0.58,0,0.82}
\definecolor{backcolour}{rgb}{0.92,0.92,0.92}

\definecolor{ieeublue}{RGB}{24, 95, 165}
\definecolor{ieeelight}{RGB}{230, 241, 251}
\definecolor{ieeedark}{RGB}{4, 44, 83}

\newmdenv[
  topline=true,
  bottomline=true,
  rightline=true,
  leftline=true,
  linewidth=1.5pt,
  linecolor=gray!60,
  backgroundcolor=gray!30,
  innertopmargin=8pt,
  innerbottommargin=8pt,
  innerleftmargin=10pt,
  innerrightmargin=10pt,
  skipabove=10pt,
  skipbelow=10pt,
  frametitle={%
    \color{black}\sffamily\bfseries\scriptsize
    \faInfoCircle\ \ KEY TAKEAWAY   
  },
  frametitlealignment=\raggedright,
  frametitlerule=true,
  frametitlerulewidth=0pt,
  frametitlebackgroundcolor=gray!45,
  frametitleaboveskip=4pt,
  frametitlebelowskip=4pt,
]{summarybox}

\def\BibTeX{{\rm B\kern-.05em{\sc i\kern-.025em b}\kern-.08em
    T\kern-.1667em\lower.7ex\hbox{E}\kern-.125emX}}
\begin{document}

\title{LLM Assisted Verification Assertion Generation: Challenges and Future Directions}

\author{\IEEEauthorblockN{Bhabesh Mali, Chandan Karfa}
\IEEEauthorblockA{Indian Institute of Technology Guwahati, India\\
\ \{m.bhabesh, ckarfa\}@iitg.ac.in}}

\maketitle

\begin{abstract}
Assertion-based Verification (ABV) plays a critical role in the Design Verification (DV) process. However, ABV requires substantial manual effort in generating assertion from specification by verification engineers, making it a time-consuming stage in the chip design flow. With the recent development of Large Language Models (LLMs), researchers have started exploring their use as an assistance in the ABV process, particularly for generating SystemVerilog Assertions (SVAs) from design specification. In this paper, we provide an overview of recent works, highlighting the different methods used to generate SVAs. In particular, we investigate LLM-based SVA generation and ask a central question: \textit{How can LLM-based assertion generation be made systematic and quality-aware?} While addressing this key question, we provide \textit{Key Takeaways} at the end of each challenge, summarizing the important methodological insights, and also provide \textit{guidelines and directions} in solving those challenges that can help generate a high-quality set of assertions using LLMs. 
\end{abstract}


\section{Introduction}
Design verification is a critical stage in the chip design flow, ensuring that an implementation satisfies its intended functional specification. Among different verification methodologies, assertion-based verification (ABV) is widely adopted as it allows design intent to be expressed as formal properties and checked against the RTL implementation. Traditionally, verification engineers manually write properties from design description and encode them as SystemVerilog Assertions (SVAs), while design engineers implement the RTL using a Hardware Description Language (HDL). These SVAs specify the expected functionality of the design and verification tool verify whether the RTL implementation satisfies the specified properties. Manually writing SVAs is labour-intensive and complex, making it time-consuming and prone to human error. Therefore, researchers have now moved towards automatic assertion generation frameworks \cite{vasudevan2010goldmine, liu2011automatic}. 

The introduction of the Transformer architecture \cite{vaswani2017attention} by Google in 2017 has provided the foundation for modern Large Language Models (LLMs). This progress has motivated the use of LLMs in ABV process, particularly focusing on generating functional SVAs automatically. 
Existing frameworks use diverse strategies for LLM-based assertion generation, including structured specification processing, multi-agent pipelines for information extraction and property generation, RTL-assisted prompting, and evaluation using different quality metrics. These variations raise an important question: \textit{How can LLM-based assertion generation be made more systematic and quality-aware by effectively representing design information, including Natural Language Specifications (NLS) and RTL?}. 

In this paper, we review existing LLM-based assertion-generation works to characterize the current state of the field. Through this analysis, we identify key design choices, limitations, and requirements for developing more systematic and quality-aware LLM-based assertion-generation flows. The results and reports of the case studies performed can be found in \cite{mali2026llmassertionlessons}. The primary contributions of this paper are as follows:
\begin{itemize}
    \item We provide a structured study of recent LLM-based assertion-generation frameworks, understanding their prompt inputs, generation strategies, refinement mechanisms, and evaluation methodologies.

    \item We identify critical quality challenges in LLM-generated assertions, including signal-name inconsistency, vacuous proofs, redundancy, incomplete functional coverage, and misleading coverage inflation, and discuss their impact on formal verification reliability.

    \item We discuss quality-aware analysis mechanisms for evaluating generated assertions, including vacuity analysis, coverage-based evaluation, and cone-of-influence (COI)-based assertion importance analysis.

    \item We conducted case studies to demonstrate the impact of different approaches and identified promising directions and provided guidelines for making LLM-based assertion generation systematic and quality-aware
\end{itemize}

    
The paper is organized as follows: 
Section II provides an overview of the different challenges encountered when generating LLM-based SVAs. Section III provides guidelines and directions for resolving the various challenges. Section IV concludes the paper.

\section{Challenges in LLM Assisted Assertion Generation}
 Assertion checking can be performed either dynamically through simulation-based testbenches (dynamic ABV) or statically using formal verification tools (static ABV). In dynamic ABV, assertions are evaluated over simulation traces, whereas in static ABV, also called as Formal Property Verification (FPV), formal tools construct a state-space representation of the design and exhaustively check whether the specified properties hold over all reachable behaviours. We focus in particular on assertion-based model checking as a formal verification technique.
However, properties are traditionally written manually, and for large IPs or industry-level designs, it can take a long time to check them against the implementation. As reported by Wilson Research Group \cite{foster2024wrg}, nearly 50\% to 60\% of chip design development time is spent on the design verification cycle. Therefore, automated assertion generation can reduce the manual effort and complexity involved in design verification. In this Section, we will discuss several challenges in LLM assisted assertion generation. 

\subsection{Specifications vs RTL}
The first fundamental question that arises in LLM-based assertion generation is: \textit{Should we generate assertions from Specification or RTL or should we use both?}
\vspace{0.2cm}

\noindent \textbf{Initial Works:} Early studies primarily used RTL descriptions as the main source of information for generating assertions. M. Orenes-Vera \textit{et al}. \cite{orenesvera2023usingllmsfacilitateformal} proposed an iterative correction loop–based framework that uses RTL code to generate functional assertions. While R. Kande \textit{et al}. \cite{kande2024security} proposed an approach to generate assertions from code comments, RTL context, and example assertions using LLMs. \textit{However, RTL may not always accurately represent the intended design behaviour, as it can contain bugs or incomplete implementations.} Formal assertions are intended to check whether the RTL satisfies the specification. \textit{Hence, generating assertions directly from RTL or code comments may produce incomplete or misleading properties}. \textcolor{black}{\textbf{Case Study 1} illustrates the observation of using RTL to generate assertions.} 

\begin{figure}[htp]
\centering
\begin{tcolorbox}[
    colback=gray!5,
    colframe=black,
    width=0.47\textwidth,
    boxrule=0.7pt,
    arc=3pt,
    left=4pt,
    right=4pt,
    top=4pt,
    bottom=4pt,
    fontupper=\small
]
\textbf{Case Study 1: Generating Assertions from RTL}

\vspace{3pt}
\hrule
\vspace{5pt}

Let us consider a 4-bit ripple-carry adder (RCA) implemented using full adders (FAs). However, the design engineer mistakenly implemented the carry-out condition in FA part of the RTL as 
$C_{out} = (A \& B) \| ((B \& C_{in}) \& (A \& C_{in}))$, 
where $A$, $B$, $C_{in}$, and $C_{out}$ are the FA signals. However, the correct carry-out condition for a full adder is 
$C_{out} = (A \& B) \| (B \& C_{in}) \| (A \& C_{in})$.

\vspace{0.2cm}

If an LLM is used to generate assertions from this buggy RTL, the properties may capture the erroneous implementation rather than the intended specification, leaving the bug undetected and reducing the effectiveness of FPV. We also verified the generated properties using Cadence JasperGold against the correct RTL: out of 14 assertions, only 5 were proven. One proven property and one property with a CEx are shown below.

\vspace{0.25cm}

\begin{tcolorbox}[
    colback=white,
    colframe=black,
    boxrule=0.5pt,
    arc=2pt,
    left=3pt,
    right=3pt,
    top=3pt,
    bottom=3pt
]

\begin{minipage}[t]{0.47\linewidth}
\centering
\textbf{Proven}

\vspace{3pt}
{\ttfamily\scriptsize
\begin{tabular}{l}
FA\_sum\_A: \\
assert property ( \\
\quad sum == \\
\quad (a \string^ b \string^ cin));
\end{tabular}
}
\end{minipage}
\hfill
\begin{minipage}[t]{0.47\linewidth}
\centering
\textbf{Property with CEx}

\vspace{3pt}

{\ttfamily\scriptsize
\begin{tabular}{l}
FA\_cout\_A: \\
assert property ( \\
\quad cout == ((a\&b)| \\
\quad ((b\&cin)\& (a\&cin))));
\end{tabular}
}
\end{minipage}

\end{tcolorbox}

\end{tcolorbox}
\label{fig:promptsignal}
\end{figure}

\noindent \textbf{Recent Works:}
To address this concern, the researchers moved toward specification-driven assertion generation, where specification serves as the primary source of design intent, while RTL is used only as supporting information for signal mapping, refinement, or verification tasks. ChIRAAG \cite{mali2024chiraag} is one of the first LLM-based frameworks for generating SVAs from NLS and evaluating them using RTL and a formal verification tool. 
Then, LAAG-RV \cite{maddala2024laag} introduced the assistance of RTL to preserve consistent signal names between SVAs and the implementation. However, its signal-alignment strategy was not sufficiently strict, leaving scope for further improvement. These works initiated a broader shift toward specification-driven SVA generation, where different frameworks use NLS, structured specifications, waveform descriptions, or knowledge representations to generate formal properties, taking RTL implementation as an assistance. More recent works in this direction include \cite{yan2025assertllm, bai2025assertionforge, gupta2025sangam}, which are explained in Section III (B). \textcolor{black}{\textbf{Case Study 2} illustrates the signal name inconsistency problem while generating SVAs.}


\begin{figure}[htp]
\centering
\begin{tcolorbox}[
    colback=gray!5,
    colframe=black,
    width=0.47\textwidth,
    boxrule=0.7pt,
    arc=3pt,
    left=4pt,
    right=4pt,
    top=4pt,
    bottom=4pt,
    fontupper=\small
]
\textbf{Case Study 2: Signal Name Inconsistency Problem}

\vspace{3pt}
\hrule
\vspace{5pt}

To showcase signal name inconsistency, we take the correct implementation of RCA from Case Study 1. An LLM may generate assertions based on signals present in the specification. Hence, those signals will not be part of the RTL.

Among the 20 assertions generated from the RCA specification, four reference a specification-only signal that is absent from the RTL implementation. Also, these four assertions do not correspond to any implemented RTL behavior and are therefore functionally invalid with respect to the RTL. One such assertion is shown below. The signal \texttt{tmp} in \textbf{Assertion 1} appears only in the specification and is missing from the RTL.

\vspace{0.2cm}

\begin{tcolorbox}[
    colback=white,
    colframe=black,
    boxrule=0.5pt,
    arc=2pt,
    left=4pt,
    right=4pt,
    top=4pt,
    bottom=4pt
]
\centering
\textbf{Assertion 1}

\vspace{3pt}
{\ttfamily\scriptsize
\begin{tabular}{l}
vacuous\_wrong\_sum\_bit0\_A: \\
assert property ( \\
\quad !tmp || (sum[0] ==  \string~(a\_in[0] \string^ b\_in[0] \string^ cin)));
\end{tabular}
}

\end{tcolorbox}

\end{tcolorbox}
\label{fig:promptsignal}
\end{figure}

\vspace{-0.2cm}

\begin{summarybox}
  \textit{\small
  \begin{itemize}
      \item SVAs should be derived from the design specification that defines the intended design behaviour
      \item Signal mapping between the specification and the implementation is crucial in LLM-based assertion generation. Precise mapping must be available during the assertion generation process.
  \end{itemize}
  }
\end{summarybox}


\subsection{Ambiguity in Natural Language Specification}
The second fundamental question is: \textit{How can design specifications be represented in a structured form to reduce ambiguity and preserve functional intent consistently during LLM-based SVA generation?}

As discussed, the recent frameworks use design specifications as the primary source for generating SVAs. However, design specifications can confuse an LLM when generating the complete set of assertions, as they are sometimes ambiguous and incomplete. \textit{This becomes a challenge for researchers to provide a uniform specification to the LLM to generate the assertions.} \textcolor{black}{\textbf{Case Study 3} presents observations on the impact of ambiguity in NLS.}

\begin{figure}[htp]
\centering
\begin{tcolorbox}[
    colback=gray!5,
    colframe=black,
    width=0.47\textwidth,
    boxrule=0.7pt,
    arc=3pt,
    left=4pt,
    right=4pt,
    top=4pt,
    bottom=4pt,
    fontupper=\small
]
\textbf{Case Study 3: Ambiguity in NLS}

\vspace{3pt}
\hrule
\vspace{5pt}
To demonstrate the problem of ambiguous NLS, we take the RCA specification from Case Study 1 and generate two sets of ambiguous assertions using the following prompt:

\vspace{0.2cm}

\begin{tcolorbox}[
    colback=white,
    colframe=black,
    boxrule=0.5pt,
    arc=2pt,
    left=4pt,
    right=4pt,
    top=4pt,
    bottom=4pt
]

\textbf{System Prompt}

\vspace{3pt}
Rewrite the attached design specification into two ambiguous versions: one moderately ambiguous and one highly ambiguous. Keep the same design idea, but make the details less clear by weakening exact behaviour, signal relationships, bit-level rules, carry behaviour, output mapping, and examples. Do not add new functionality or contradict the original specification.

\end{tcolorbox}

\vspace{0.2cm}
We generated two ambiguous specification sets, \textbf{Ambiguity Set 1 (AS1)} and \textbf{Ambiguity Set 1 (AS2)}, with medium and high levels of ambiguity, respectively, of generalized RCA design details. LLM-based SVA generation produced 15 assertions from AS1 and 8 assertions from AS2. AS2 also generated assertions with completely different signal names from RTL, while adding a new signal to it. Although the generated assertions are useful for validating the input/output functionality of the 4-bit RCA, they are insufficient to confirm the correctness of the internal FA-based ripple-carry implementation. In particular, both AS1 and AS2 do not directly verify the internal ripple carry structure. While all assertions are proven during FPV, the formal coverage remains low: 15.91\% for AS1 and 7.50\% for AS2.

\end{tcolorbox}
\label{fig:promptsignal}
\end{figure}

\noindent \textbf{Recent Works:} Initial works such as ChIRAAG \cite{mali2024chiraag} and LAAG-RV \cite{maddala2024laag} involve extracting the important information from design specifications and transforming them into JSON format. AssertLLM \cite{yan2025assertllm} and SANGAM \cite{gupta2025sangam} extract signal-wise information from the design specification, where each signal is associated with structured descriptions such as its definition, functionality, interconnections, additional constraints, and related signals. However, SANGAM generates assertions using the Monte Carlo Tree Self-refine (MCTSr) method, forming a tree of depth four where each node in the tree represents a set of assertions. Building on such structured extraction, recent works have explored richer specification representations to provide LLMs with more complete and precise design-intent information, reducing the ambiguity. AssertionForge \cite{bai2025assertionforge} structured their specification into a Knowledge Graph (KG), containing ports, registers, modules, etc., as node entities and the edges as relationships between them, if they exist. However, they have also used the RTL to further enhance the KG and provide completeness to generate a better set of assertions. While AssertCoder \cite{tian2025assertcoder} generates SVAs from multimodal design specifications by first segmenting and classifying the specification contents into different modalities, including text, diagrams, formulas, and tables. It uses dedicated LLM-based analyzers for each modality to extract structured semantic information, which is then merged into a unified specification representation to support the SVA generation task. \textcolor{black} {\textbf{Case Study 4} analyzes the impact of different NLS representations on assertion generation results.}

\begin{figure}[htp]
\centering
\begin{tcolorbox}[
    colback=gray!5,
    colframe=black,
    width=0.47\textwidth,
    boxrule=0.7pt,
    arc=3pt,
    left=4pt,
    right=4pt,
    top=4pt,
    bottom=4pt,
    fontupper=\small
]
\textbf{Case Study 4: Different Representation of NLS}

\vspace{3pt}
\hrule
\vspace{5pt}

To analyse the impact of NLS representation on generating a set of assertions, we study three recent SOTA LLM-based assertion generation frameworks: SANGAM \cite{gupta2025sangam}, AssertLLM \cite{yan2025assertllm}, and AssertionForge \cite{bai2025assertionforge}. The analysis is performed using the \texttt{UART} design as a representative case study.
\vspace{0.2cm}

\begin{tcolorbox}[
    colback=white,
    colframe=black,
    boxrule=0.5pt,
    arc=2pt,
    left=4pt,
    right=4pt,
    top=4pt,
    bottom=4pt
]

\textbf{Comparison Table}

\vspace{2pt}

\begin{center}
\scriptsize
\setlength{\tabcolsep}{2pt}
\renewcommand{\arraystretch}{1.15}
\begin{tabularx}{\linewidth}{p{1.55cm}|X|c|c}
\hline
\textbf{Framework} &
\textbf{NLS Representation} &
\textbf{\# Assert.} &
\textbf{FPV Pass} \\
\hline
AssertLLM (AL) &
Signal-wise structured form with RAG-based enhancement &
249 &
29 \\
\hline
SANGAM (SG) &
Signal-wise structured form with Monte Carlo Tree Self-Refine and assertion rectification loop &
276 &
107 \\
\hline
AssertionForge (AF) &
Knowledge Graph (KG)-based representation using nodes for I/O signals, ports, registers, etc., and depth-based KG traversal &
253 &
27 \\
\hline
\end{tabularx}
\end{center}

\end{tcolorbox}

\vspace{0.2cm}
We observe that SANGAM \cite{gupta2025sangam} has the highest number of proven assertions, mainly due to its improvement loop, which refines assertions using feedback from formal-tool logs. However, AssertionForge \cite{bai2025assertionforge} achieves higher functional coverage, with 88.82\%, compared to 66.78\% for AssertLLM \cite{yan2025assertllm}. This indicates that a higher number of proven assertions does not necessarily imply higher functional coverage. SANGAM \cite{gupta2025sangam} reported a timeout issue, and hence, the functional coverage couldn't be reported.

\end{tcolorbox}
\label{fig:promptsignal}
\end{figure}

Despite these efforts, there is still no standardized approach for preprocessing and representing design specifications for assertion generation. Directly using raw specifications as input to LLMs may lead to incomplete or inaccurate assertions due to ambiguity, incompleteness, and heterogeneous specification formats. 
\textit{This motivates the need for a structured specification representation that can support robust assertion generation while using RTL only for necessary implementation-level details.}

\begin{summarybox}
  \textit{\small
    Design Specification should be structured into some precise/uniform representation while keeping consistent signal names wrt implementation, such that we can provide robust information to the LLM to generate SVAs.
  }
\end{summarybox}

\subsection{Assertion Generation and Verification}
The next step after specification processing is the assertion generation. Different frameworks have used different types of LLM(s) and techniques to generate the set of assertion and rectify them.
\vspace{0.15cm}

\noindent \textbf{One-shot Methods:}
AssertLLM \cite{yan2025assertllm}, AssertionForge \cite{bai2025assertionforge} and security assertion generation framework by Kande \textit{et.al} \cite{kande2024security} do not include an explicit FPV-guided rectification loop for rectifying generated SVAs, i.e., the methods are one shot. Kande \textit{et.al} \cite{kande2024security} evaluated the generated assertion set using simulation testbenches, which they described as exhaustive. While  AssertLLM \cite{yan2025assertllm} and AssertionForge \cite{bai2025assertionforge} have used FPV tool to evaluate generated SVAs by checking syntax correctness and property pass/fail status. 
\vspace{0.15cm}

\noindent\textbf{Rectification Loop-based Methods}: An early work \cite{orenesvera2023usingllmsfacilitateformal} generated assertions by prompting LLMs with RTL descriptions. After generation, the assertions were evaluated using an FPV tool to determine whether they were proven or produced counterexamples (CEx). In the case of a CEx, the verification results were manually inspected, and the prompt rules were refined to guide the LLM toward generating corrected assertions. While, testbench-based evaluation and refinement of LLM-generated assertions has been adopted in works such as ChIRAAG \cite{mali2024chiraag} and LAAG-RV \cite{maddala2024laag}. These frameworks check the generated assertions for syntax errors and evaluate whether the assertions pass during verification. If an assertion fails or cannot be resolved after a predefined number of refinement attempts, human intervention is required. While \cite{gupta2025sangam} is another work that uses an assertion rectification loop, they have used a formal tool and can therefore perform an exhaustive evaluation. 

AssertCoder \cite{tian2025assertcoder}, on the other hand, evaluates initially generated LLM assertions through mutation injection and model checking. Specifically, it creates assertion-mutation pairs by injecting mutations into the design and checking whether the generated assertions fail under the mutated behavior. The framework further defines a \textbf{Mutation Score}, which measures the number of unique mutations detected by an assertion. A mutation score of zero indicates that the assertion has low bug-detection capability and should be removed. In addition, AssertCoder defines an aggregate metric, \textbf{Mutation Detection Rate (MDR)}, which is used to decide whether the SVA should be regenerated. \textcolor{black}{\textbf{Case Study 5} reports an observation while using one-shot and loop-based assertion generation methods.}

\begin{figure}[htp]
\centering
\begin{tcolorbox}[
    colback=gray!5,
    colframe=black,
    width=0.47\textwidth,
    boxrule=0.7pt,
    arc=3pt,
    left=4pt,
    right=4pt,
    top=4pt,
    bottom=4pt,
    fontupper=\small
]
\textbf{Case Study 5: One-shot vs Loop-based Method}

\vspace{3pt}
\hrule
\vspace{5pt}
As shown in \textbf{Case Study 4}, the number of assertions generated and proven in one-shot methods (AssertLLM and AssertionForge), and rectification loop-based methods (SANGAM). We can observe that only 11.6\% and 10.6\% of the generated assertion in AssertLLM \cite{yan2025assertllm} and AssertionForge \cite{bai2025assertionforge} are syntactically and semantically correct. While, in case of SANGAM \cite{gupta2025sangam}, because of the rectification loop, around 38.7\% of the generated assertion are proven.

\end{tcolorbox}
\end{figure}

\vspace{-0.5cm}

\begin{summarybox}
  \textit{\small
    An FPV tool should be used to validate the generated assertions by identifying whether they are proven, fail due to syntactic or semantic errors, or produce CEx.
  }
\end{summarybox}

\subsection{Assertion Evaluation}
After LLM-guided assertion generation and assertion rectification, the resulting assertion set must be evaluated to determine the quality and effectiveness of the generated assertions.

\noindent\textbf{Existing Works:} As such SANGAM \cite{gupta2025sangam} uses coverage metrics such as \textit{Branch Coverage, Property Coverage} and \textit{Toggle Coverage} to check how much of the design implementation is actually been covered by the LLM-generated assertions. Similarly, AssertionForge \cite{bai2025assertionforge} used \textit{Statement, Branch, Functional} and \textit{Toggle} coverage metrics to check the quality of generated assertions. While \cite{mali2024chiraag, maddala2024laag, yan2025assertllm, tian2025assertcoder} haven't used any coverage metrics, rather relied on the whether the assertions are proven without any CEx. Also, existing LLM-based assertion generation frameworks have not taken strong measures to address the signal mapping process. 

While prior works \cite{maddala2024laag, gupta2025sangam, bai2025assertionforge} introduce preliminary mechanisms to improve consistency between generated assertions and RTL signals, their prompt-based strategies do not appear sufficiently robust. The main concern is that signal inconsistency between the RTL implementation and the generated assertions can lead to vacuous proofs. In particular, if an antecedent signal used in an assertion is not present in the RTL and the LLM is hallucinated during assertion generation from the NLS, the antecedent may never be activated, causing the assertion to be proven vacuously within a few cycles. \textbf{This can falsely inflate the reported coverage metrics.} 
\textit{This motivates the need for new analysis metrics and mitigation strategies to reduce such issues.} \textcolor{black}{\textbf{Case Study 6} highlights the insufficiency of existing evaluation metrics and the need for vacuity analysis.}
\vspace{-0.15cm}
\begin{figure}[htp]
\centering
\begin{tcolorbox}[
    colback=gray!5,
    colframe=black,
    width=0.47\textwidth,
    boxrule=0.7pt,
    arc=3pt,
    left=4pt,
    right=4pt,
    top=4pt,
    bottom=4pt,
    fontupper=\small
]

\textbf{Case Study 6: Evaluation Metrics insufficiency and Vacuity}

\vspace{3pt}
\hrule
\vspace{5pt}

Here we demonstrate how signal-name inconsistency can affect assertion evaluation and lead to misleading conclusions about the quality of the generated assertion set. We consider the \texttt{i2c} \cite{yan2025assertllm} and \texttt{GA\_68} \cite{wan2026fixme} designs using two frameworks: SANGAM \cite{gupta2025sangam} and AssertionForge \cite{bai2025assertionforge}. We observe that both frameworks do not strongly restrict the signal-mapping process. As a result, several signal names generated in the SVAs are not present in the RTL, so the antecedent may never be activated, causing the assertion to be proven vacuously and falsely inflating the coverage results. The coverage is high primarily due to vacuous assertions as shown in the table below.

\vspace{0.2cm}

\begin{tcolorbox}[
    colback=white,
    colframe=black,
    boxrule=0.5pt,
    arc=2pt,
    left=3pt,
    right=3pt,
    top=4pt,
    bottom=4pt
]

\textbf{Vacuity and Coverage Comparison}

\vspace{3pt}

\begin{center}
\scriptsize
\setlength{\tabcolsep}{1.8pt}
\renewcommand{\arraystretch}{1.2}
\begin{tabular}{llcccccc}
\hline
\textbf{Design} & \textbf{FW} & \textbf{Raw} & \textbf{Vac.} & \textbf{Vac.} & \textbf{FC} & \textbf{SC} & \textbf{CC} \\
 &  & \textbf{Assert.} & \textbf{Assert.} & \textbf{(\%)} & \textbf{(\%)} & \textbf{(\%)} & \textbf{(\%)}\\
\hline
\texttt{i2c} & SG & 245 & \textbf{\textcolor{red}{72}} & 29.38 & 85.87 & 89.06 & 92.16 \\
             & AF & 256 & \textbf{\textcolor{red}{58}} & 22.65 & 73.42 & 92.84 & 68.98 \\
\texttt{GA\_68} & SG & 180 & \textbf{\textcolor{red}{39}} & 21.66 & 73.52 & 90.19 & 66.06 \\
                & AF & 224 & \textbf{\textcolor{red}{75}} & 35.88 & 70.33 & 86.15 & 64.95 \\
\hline
\end{tabular}
\end{center}

\vspace{1mm}
{\footnotesize
\textit{FW: Framework, Vac.: Vacuous, FC: Formal Coverage, SC: Stimuli Coverage, CC: Checker Coverage, SG: SANGAM, AF: AssertionForge.}
}

\end{tcolorbox}

\end{tcolorbox}
\label{fig:vacuity_coverage_case_study}
\end{figure}

\begin{summarybox}
  \textit{\small
    Coverage metrics such as Formal Coverage, Branch Coverage etc., are not sufficient to define the quality of the metrics.
  }
\end{summarybox}

\subsection{Quality of the Generated Assertion}
Existing SOTA LLM-based assertion generation frameworks, such as SANGAM \cite{gupta2025sangam}, AssertLLM \cite{yan2025assertllm}, and AssertionForge \cite{bai2025assertionforge}, provide limited mechanisms for rigorously evaluating the quality of the final generated assertion set. In particular, they do not strongly analyze: \textbf{\textit{(1)}} whether the final assertion set contains redundant assertions, and \textbf{\textit{(2)}} whether some assertions are logically dependent on, or weaker than, other assertions in the set. Identifying such redundancy and dependency is important, as removing equivalent or weaker assertions can reduce the size of the assertion set without significantly degrading evaluation metrics. \textit{This motivates the need for additional assertion-reduction techniques that retain only a compact and high-quality set of assertions.}

\subsection{Handling Long-Context Design Specifications}
LLMs have a limited context window, which restricts the amount of specification, RTL, and auxiliary information that can be provided in a single prompt \cite{an2024make}. Directly providing such documents to an LLM for functional assertion generation may produce unstable, incomplete, or inconsistent assertion sets. This issue becomes more critical for industrial IPs/designs, where both the specifications and RTL implementations are typically larger, more complex, and richer in design intent. \textit{These observations motivate the need to organize, abstract, and transform specification and RTL information into a structured representation that preserves functional behavior while enabling scalable LLM-based assertion generation.} 

\begin{summarybox}
  \textit{\small
   Generating SVA from large design specification and RTL is a challenging task as the token limit of LLM(s) are limited.
  }
\end{summarybox}

\section{Guidelines and Future Directions}

\begin{figure}[!h]
    \centering
    \includegraphics[width=0.42\textwidth]{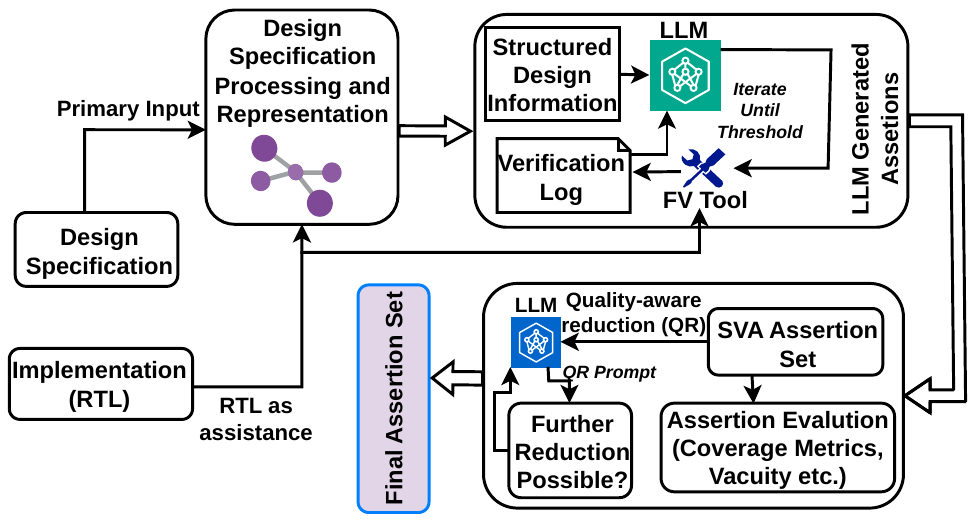}
    \caption{Overall Generalized Steps for LLM-based Assertion Generation}
    \label{fig:overall}
\end{figure}

With respect to the existing works and experiments performed, we provide some guidelines for the future researchers to generate SVA using LLM, while integrating them into their framework. Fig. \ref{fig:overall} shows an overview of the different steps involved in LLM-based SVA generation.

\subsection{\textbf{SVA should be generated from design specification}}
In LLM-based SVA generation, the primary source of input should always be the design specification. However, RTL can be used as an assistance in certain steps such as signal mapping, FPV etc. The RTL functionality should never be used to generate SVA assertions.
\subsection{\textbf{Mandatory addition of precise signal mapping step in the framework}}
 In LLM-based SVA generation, signal mapping is a critical preprocessing step. In practical verification settings, design and verification engineers typically agree on the input, output, wire, and register names before implementing the RTL and writing the corresponding properties. Therefore, before generating assertions, the RTL signal names (input/output ports/ wires/registers) should be extracted and provided to the LLM with strict instructions to use only valid RTL signals and avoid introducing any undeclared or specification-only signal names. Any generated signal that does not exactly match an RTL signal should be mapped to the closest valid RTL signal before formal verification.

 \subsection{\textbf{Resolving Ambiguity in NLS with Advanced representation}}
To reduce ambiguity in NLS, a KG-based specification representation can serve as an effective preprocessing step, where signals, such as inputs, outputs, wires, and registers, and design elements are modeled as nodes, while their interactions are modeled as edges. The RTL can also be used as an assistance to add missing or incomplete signal-level information; however, it should not be treated as the primary source of design intent.

\subsection{\textbf{Beyond Coverage Metrics: Evaluating LLM-based SVAs}}
As we have discussed and shown in \textbf{Case Study 6}, the coverage metrics are insufficient. In this regard, we discuss several possible directions below: 

\textbf{(1)} Beyond coverage metrics, another important quality indicator is the \textit{\textbf{Vacuity Metric}}. This metric helps distinguish assertions that are genuinely proven by the formal tool from those that are proven vacuously due to issues such as inactivated antecedents, over-constrained environments, or signal name mismatches etc. FV tools, such as Cadence JasperGold, report vacuous and non-vacuous properties, which can be used to quantify the extent of vacuous proofs in the generated assertion set. Furthermore, the vacuous assertion set can be passed to an LLM-based refinement framework to identify and mitigate the underlying causes of vacuity, including signal mismatches between the generated assertions and the RTL implementation at that level. 

\textbf{(2)} 
Another important aspect to investigate is the \textbf{\textit{Cone-of-Influence (COI)}} of each assertion (wrt signals). By identifying unique COIs, one can analyze the functional impact of individual assertions on the design. Assertions with broader or more distinct COIs may be prioritized first, and their contribution can be evaluated using coverage metrics. Subsequently, additional assertions whose COIs cover lower-priority or previously uncovered design regions can be incrementally added, while observing the corresponding changes in coverage. This process can help identify the most influential assertions that fully or partially capture the functional behavior of the design.

\subsection{\textbf{Redundant and weak assertions should be removed from the final assertion set.}}

After generating the final assertion set, additional steps should be taken to analyze redundant and weak assertions. Since FPV can be time-consuming, reducing the assertion set while preserving its functional intent can help improve verification efficiency. One possible direction is to perform redundancy and strong/weak analysis among the generated assertions. Redundant assertions can be identified through equivalence checking between assertion pairs; if two assertions are equivalent, one of them can be removed \cite{10.5555/1208179}. After eliminating redundant assertions, strong/weak relationships can be analyzed. If one assertion is stronger than another, the weaker assertion can be removed, as it is subsumed by the stronger one \cite{10.5555/1208179}.

\subsection{\textbf{Inclusion of Rectification Loop}}
 Current LLM-based assertion generation methods remain prone to errors, as LLMs may misinterpret prompts or generate incorrect responses. Therefore, instead of using one-shot methods to generate assertions, future frameworks should use a formal log-guided rectification loop. In such a loop, an FPV tool verifies the generated SVAs against the RTL implementation and produces logs that capture syntactic errors, semantic errors, and counterexamples. These logs, together with the original assertions, can then be provided to an LLM for refinement, enabling the generation of a more correct assertion set.

\subsection{\textbf{Long-Context problem should be handled with advanced representation of NLS and RTL}}

To address this challenge, one promising direction is to structure and utilize large-scale specification and RTL information through signal-wise representations. Tools such as Microsoft GraphRAG \cite{microsoft_graphrag_docs}, LlamaIndex \cite{llamaindex_llamaparse_docs} etc., can be used to construct a structured representation, e.g., a knowledge graph, of both the specification and RTL in a \textbf{signal-wise} manner. In such settings, generating all possible functional assertions in a single step may be ineffective. Instead, assertions can be generated incrementally at the signal level. For example, when using a KG, the relevant specification information associated with each target signal can be retrieved and organized into a signal-specific context. This context can then be provided to an LLM for assertion generation, avoiding the need to supply the entire design information in a single prompt.

\section{Conclusion}
In this paper, we present a detailed analysis of recent LLM-based SVA-generation frameworks. We examine how existing methods use RTL implementations, NLS, or both as input sources for assertion generation. We further analyze different design-representation strategies for extracting assertion-relevant information and discuss the evaluation metrics used to assess the quality of LLM-generated SVAs. Furthermore, we provide \textit{Key Takeaways} and \textit{guidelines} which can serve as practical direction for the future researchers for generating high-quality assertion sets.

\bibliographystyle{IEEEtran}
\vspace{-0.1 cm}
\bibliography{ref.bib}

\end{document}